\documentclass[11pt,twoside]{article}
\usepackage{asp2010}

\usepackage{graphicx}
\resetcounters

\begin{document}

\title{Symmetric vs. asymmetric planetary nebulae:  morphology and chemical abundances}
\author{W. J. Maciel and R. D. D. Costa
\affil{Astronomy Department, University of S\~ao Paulo, Rua do Mat\~ao 1226, 
05508-090 S\~ao Paulo SP, Brazil}}

\begin{abstract}
We analyse a large sample of galactic planetary nebulae based on their chemical composition and 
morphology. A recent morphological classification system is adopted, and several elements are 
considered, namely He, N, O, S, Ar, Ne, and C in order to investigate the correlations involving 
these elements and the different PN types. Special emphasis is given to the differences between 
symmetric (round or elliptical) nebulae and those that present some degree of asymmetry 
(bipolars or bipolar core objects). The results are compared with previous findings both for PN  in 
the Galaxy and in the Magellanic Clouds.

\end{abstract}


\section{Introduction}

Planetary nebulae (PN) may have different morphological aspects, as made evident by many classification 
schemes proposed since the pioneering work of \cite{curtis}. Different morphologies are probably 
related to different physical origins, which may include the effects of different stellar masses, binary 
evolution, magnetic fields, etc. As a consequence, the observed variety of PN morphologies is associated 
with different physical properties, which include their chemical abundances. In this work, we consider a 
large sample of PN in the Galaxy for which high resolution optical images and accurate abundances are 
available, and investigate the relationship between the morphological characteristics and the chemical 
abundances, especially concerning symmetric and asymmetric nebulae.

\section{Morphology, abundances, and stellar evolution}

Generally speaking, the observed morphologies of PN can be interpreted either as the result of projection 
effects on a basic tridimensional structure or as a result of different physical processes affecting the 
PN origin and evolution. Recent work seems to favour the second approach, so that we will adopt the scheme 
proposed by  \cite{scs} and later updates, and look for correlations involving 
the morphologies and the nebular chemical abundances.  The PN types are: Round (R), Elliptical (E), Bipolar 
(B), Bipolar-core  (BC), and Point symmetric (P). Although some nebulae do not fit exactly in this scheme, 
most objects for which detailed high-resolution images are available can be classified in one of these types. 

According to some previous results on the relationship between the morphology  and the evolution of 
intermediate mass stars (IMS) that originate the planetary nebulae, bipolar nebulae seem to be formed by
relatively massive progenitor stars, those closer to the high mass limit of the IMS. This is consistent 
with some characteristics of these nebulae, such as a lower scale height relative to the galactic plane 
or higher helium and nitrogen abundances. The He and N enhancements can be expected in bipolar nebulae if 
they are originated from the higher mass IMS, for which the dredge-up and Hot Bottom Burning (HBB) processes 
lead to a reduction in the C abundances, while the He/H and N/O ratios are enhanced. On the other hand, most 
elliptical and round PN seem to be formed by lower mass stars, those closer to one solar mass on the 
main sequence. 

\section{The sample}

In this project we take into account elements  that are manufactured by the PN progenitor stars 
(He, N, and C), as well as some elements that are not (O, Ne, S, and Ar). We have inspected many high 
resolution images and compared our results with previous classifications in the literature.
The results reported here refer to two samples  of galactic PN, which are probably the largest samples 
for which {\it both} the morphology {\it and} chemical abundances have been analysed: Sample A,
with 234 nebulae, includes PN in the Milky Way disk, as recently analysed by \cite{mcu} [see 
also  \cite{mci10} and references therein], and  Sample B, which includes the PN of Sample A, 
supplemented by several galactic objetcs for which recent abundances have been determined in the literature. 
This sample is in principle less accurate than Sample A, but is considerably larger, presently including  
372 nebulae. 

\section{Results}
The main results are shown in Tables 1 and 2, and in Figure 1. For the sake of completeness, we have included 
the less accurate C/H  data from the literature, in view of its importance in reflecting the nucleosynthesis 
of PN progenitor stars. From Table 1 we note that symmetrical PN (R, E) are more frequent in the 
Milky Way than asymmetric (B, BC, P) objects, both for Sample A and B. From these samples, bipolars correspond 
to approximately 25\% of all galactic nebulae. As a comparison, the corresponding numbers for the Magellanic 
Clouds from \cite{stanghellini} are 46\% (S) and 54\% (A) for the LMC and 64\% (S) and 36\% (A) 
for the SMC. However, it should be mentioned that the samples considered are not complete, which also applies 
to the Magellanic Clouds.

\begin{table}[!ht]
\caption{Distribution of the galactic PN according to the morphological types}
\smallskip
\begin{center}
\small
\begin{tabular}{lrr}
\tableline
\noalign{\smallskip}
Type     & Sample A  & Sample B \\
\noalign{\smallskip}
\tableline
\noalign{\smallskip}
Round = R           & 24.8 & 21.2 \\
Elliptical = E      & 31.2 & 30.1 \\
Bipolar = B         & 16.2 & 16.7 \\
Bipolae Core = BC   &  6.4 &  7.0 \\
Point-symmetric = P  &  3.0 &  1.9 \\
Unclassified = U    & 18.4 & 23.1 \\
\tableline
\noalign{\smallskip}
Symmetric = S       & 56.0 & 51.3 \\
Asymmetric = A      & 25.6 & 25.5 \\
\noalign{\smallskip}
\tableline
\end{tabular}
\end{center}
\end{table}
\begin{table}[!ht]
\caption{Average abundances of symmetric and asymmetric galactic PN}
\smallskip
\begin{center}
\small
\begin{tabular}{lcc}
\tableline
\noalign{\smallskip}
   &  Sample A & Sample B \\
\noalign{\smallskip}
\tableline
\noalign{\smallskip}
He/H: \ \ All&  $0.114 \pm  0.024$ & $0.113 \pm 0.025$ \\
Symmetric    &  $0.107 \pm  0.018$ & $0.104 \pm 0.019$ \\
Asymmetric   &  $0.129 \pm  0.027$ & $0.132 \pm 0.027$ \\
\noalign{\smallskip}
O/H: \ \ All &  $8.63  \pm  0.26$  & $8.59  \pm 0.28$  \\
Symmetric    &  $8.60  \pm  0.26$  & $8.55  \pm 0.28$  \\
Asymmetric   &  $8.66  \pm  0.27$  & $8.64  \pm 0.25$  \\
\noalign{\smallskip}
S/H:\ \ All  &  $6.88  \pm  0.35$  & $6.87  \pm 0.37$  \\
Symmetric    &  $6.82  \pm  0.29$  & $6.82  \pm 0.24$  \\
Asymmetric   &  $6.92  \pm  0.40$  & $6.92  \pm 0.38$  \\
\noalign{\smallskip}
Ar/H:\ \ All &  $6.39  \pm  0.31$  & $6.37  \pm 0.33$  \\
Symmetric    &  $6.33  \pm  0.29$  & $6.30  \pm 0.31$  \\
Asymmetric   &  $6.48  \pm  0.31$  & $6.47  \pm 0.32$  \\
\noalign{\smallskip}
N/H:\ \ All  &  $8.14  \pm  0.44$  & $8.10  \pm 0.46$  \\
Symmetric    &  $8.02  \pm  0.39$  & $7,94  \pm 0.43$  \\
Asymmetric   &  $8.34  \pm  0.42$  & $8.37  \pm 0.40$  \\
\noalign{\smallskip}
Ne/H:  \ All &  $7.96  \pm  0.26$  & $7.94  \pm 0.31$  \\
Symmetric    &  $7.95  \pm  0.24$  & $7.91  \pm 0.28$  \\
Asymmetric   &  $8.01  \pm  0.28$  & $8.02  \pm 0.30$  \\
\noalign{\smallskip}
C/H:  \ All  &  $8.67  \pm  0.37$  & $8.66  \pm 0.39$  \\
Symmetric    &  $8.74  \pm  0.28$  & $8.71  \pm 0.34$  \\
Asymmetric   &  $8.47  \pm  0.47$  & $8.51  \pm 0.45$  \\
\noalign{\smallskip}
\tableline
\end{tabular}
\end{center}
\end{table}
%

   \begin{figure}[!ht]
   \centering
   \includegraphics[angle=0,width=10cm]{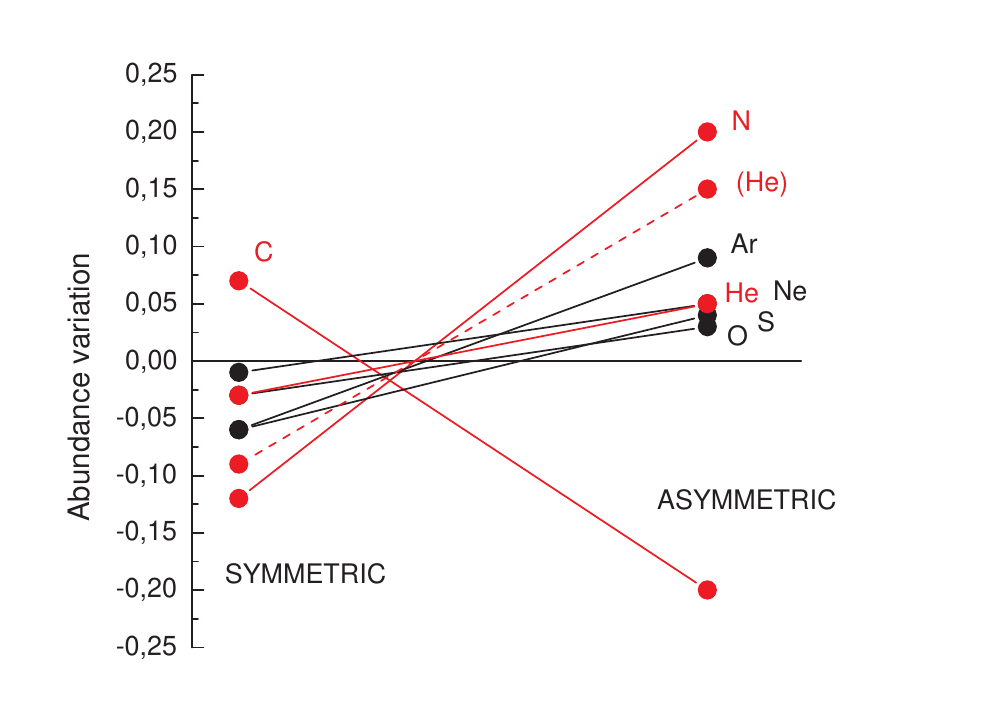}
       \caption{Abundance variations between symmetric and asymmetric PN.}
       \label{fig1}
   \end{figure}

Table 2 shows the elemental abundances defined as usual, $\epsilon$(X) = $\log$ (X/H) + 12. It can be  seen 
that both samples A and B reveal the same trends, that is (i) there is a continuous increase in the He/H 
abundances in the sequence {\it Symmetric -- All -- Asymmetric} PN, that is, asymmetric nebulae have 
systematically larger He/H abundances than symmetric objects; (ii) the same is true for the N/H and N/O ratios, 
and (iii) the inverse is true for the C/H ratio. Exactly the same trends are observed in Magellanic Cloud nebulae 
for the ratios C/H, N/H, and N/O,  according to  \cite{stanghellini}.  Therefore, these trends 
appear to hold for objects of different metallicities. These results confirm that the asymmetric nebulae are 
probably originated from the higher mass IMS, while the symmetric objects come from progenitor stars with lower masses.  

Figure 1 shows  the differences $\Delta\epsilon  = \epsilon - \epsilon$(all) for all elements for Sample A. It can 
be seen more clearly that the He and N abundances increase for asymmetric PN, while the opposite is true for C.  
A similar behaviour occurs for the elements that are not produced by the PN progenitor stars (O, S, Ne, and Ar). 
However, the abundance variations of these elements are relatively small, and probably reflect the fact the asymmetric 
nebulae are formed by more massive, younger stars, which are themselves formed out of an enriched interstellar medium. 
In the case of He, it should be recalled that the abundance uncertainties are about a factor 3 smaller than for the 
heavier elements, so that in practice the variation in the He abundances in Figure 1 could be better represented 
by the dashed line shown in the figure. These trends are in good  agreement with recent theoretical models by   
\cite{marigo} and \cite{karakas}. 

It should be noted that some bipolar nebulae do not show He or N enhancements, which may be 
explained by their origin being related to common envelope as a consequence of binary star evolution. 
The fact that there is some measurable He and N enhancements in most B or BC 
nebulae suggests that common envelope evolution is probably a secondary aspect of  PN formation. For example, in 
Sample A, 72\% of the B/BC nebulae have enhanced He/H abundances (He/H $>$ 0.120), while for 28\% of the objects 
the He/H ratio is normal, that is  He/H $<$ 0.120, suggesting that about one third of the bipolar nebulae have
common envelope origin. 

\acknowledgements This work was partially supported by FAPESP/CNPq.

\bibliography{maciel3}

\end{document}